\documentclass[preprint,showpacs,preprintnumbers,amsmath,amssymb,superscriptaddress]{revtex4-2}
\usepackage{graphicx,graphics}   % Permite a inser??o de figuras EPS, PDF, JPG ou PNG

\usepackage{graphicx,amsfonts}% Include figure files
\usepackage{epsfig}

\usepackage{dcolumn}% Align table columns on decimal point
\usepackage{bm}% bold math
\usepackage{pslatex}
\usepackage{epsfig}
\usepackage{color}
\usepackage{cancel}
\usepackage{slashed}
\usepackage{amssymb}
\usepackage{amsmath}
\usepackage{hyperref}
\usepackage[utf8x]{inputenc}
\usepackage{color}
\usepackage{multirow}
\usepackage{slashed}
\usepackage{ulem}
\usepackage{pstricks}
\usepackage[caption=false]{subfig}
\begin{document}

%\preprint{IFT-P.xx/2009}
%\preprint{ArXiv:yymm.nnnn}
\title{
Is the charged vecton boson seen by the CDF II the standard model one?}

\author{O. P. Ravinez}
\email{opereyra@uni.edu.pe -  ORCID:0000-0002-4143-8813}
\affiliation{ Facultad de Ciencias, Universidad Nacional de Ingenieria (UNI)\\
	av. Tupac Amaru s/n, Lima - R\'\i mac 15333, Per\'u}

\author{ Henry Diaz}%
\email{hdiaz@uni.edu.pe -  ORCID:0000-0001-6110-6130}
\affiliation{ Facultad de Ciencias, Universidad Nacional de Ingenieria (UNI)\\
av. Tupac Amaru s/n, Lima - R\'\i mac 15333, Per\'u}

\author{V. Pleitez}%
\email{v.pleitez@unesp.br -  ORCID: 0000-0001-5279-8438}
\affiliation{
Instituto  de F\'\i sica Te\'orica--Universidade Estadual Paulista (UNESP)\\
R. Dr. Bento Teobaldo Ferraz 271, Barra Funda\\ S\~ao Paulo - SP, 01140-070,
Brazil
}

\date{08/21/2024%08/01/2018
%07/11/15% It is always \today, today,
}
\begin{abstract}
After the CDF II results, we can ask ourselves, have they observed the vector boson $W$ of the standard model? or have they discovered the effect of new physics?
We show that, in electroweak models with $SU(2)_L\otimes SU(2)_R\otimes U(1)_{B-L}$ gauge symmetry, it is possible to accommodate, at tree level, the value of the mass of the charged vector boson obtained by the CDF-II. A shift in the mass of the lightest neutral vector boson is also predicted in some versions of the model.
\end{abstract}

\pacs{12.60.Fr; %Extensions of electroweak Higgs sector
12.15.-y }
\maketitle
%\newpage
\section{Introduction}
\label{sec:intro}

Recently the CDF II collaboration has obtained a new value for the mass of the charged vector boson SM's $W$~\cite{CDF:2022hxs}:
\begin{equation}
M_W\vert_{CDF-II} = 80,433.5 \pm 9.4 \;\textrm{MeV},	
\label{cdf}
\end{equation}
which is not consistent with the standard model (SM)'s prediction with an accuracy of 7$\sigma$. Moreover, the LEP-Tevatron combination gives the value $M_W=80,385\pm15$ MeV, in concordance with the SM theoretical calculus. {On the other hand the value given in in Eq.~(\ref{cdf}) is also in disagreement with other experimental data~\cite{Workman:2022ynf,atlas2023}. If the CDF II is correct, it could imply new physics beyond the standard model (BSM) which may be related only to the gauge bosons in the new context, certainly more measurements are needed. 

In fact in the context of new physics beyond the SM, it is possible that the vector bosons, $W$ and $Z$, which were discovered in the 
eighties are not exactly those of the standard model. It is worth remembering that the vector bosons $W$ and $Z$ were discovered at CERN in 1983 assuming that they were the only particles that could produce charged leptons and their respective neutrinos. The existence of new charged particles that can produce these leptons, would certainly imply that the $W$ mass cannot be \textit{exactly} the same obtained if only the SM is assumed. 
In the context of the SM the $W$ depends on the imput parameter as $M_Z,\alpha, G_F$ and also on the Higgs mass $m_H$. In fact, most extensions of the standard model new inputs have to be introduced and, the $W$ mass would be different from the SM value (hope within the experimental error) at some level data's precision.
It is well known that any physics beyond the standard model (BSM) can also produce these leptons with a single charged vectors (or scalar) not included in the SM which may give new contributions to lepton production will imply that the mass of $W$ is slightly different from that obtained in the SM context (this may happen with the neutral vector boson, $Z$, as well). In this case, it would be very possible that the particles that correspond to the SM continue to have the same old values while the mass obtained from CDF-II should be interpreted in the context of new physics BSM~\cite{Barela:2023eoj}. 

Some extensions of the SM have been proposed in the literature to explain the difference of the value in the $W$ mass between the SM prediction and that measured for the CDF II. For doing this, the representation content of the SM model is modified, in several ways to explain the mass of $W$. 
For instance, an extension of the SM in which a second scalar doublet carrying lepton number $L=-2$, and a singlet with $L=+1$ can be added for this purpose. Since this is not a minimal version of the model then it is mandatory to verify if the prediction of the $W$’s mass is consistent with other processes like the electroweak precision test, neutrino physics, rare decays, and collider’s results. The new charged scalar must have a mass below a few TeV. See Ref.~\cite{Batra:2022arl} for details.
The same happens with the extension of the SM in which a singlet and a triplet scalars are added, the triplet carries on $Y=0$ and $L=-3$~\cite{Popov:2022ldh}.  

It is also possible to solve the $W$ mass shift by adding to the SM minimal representation content a scalar triplet with $Y=0$ and a vector-like lepton that can be an $SU(2)_L$ doublet with $Y=-1$, or a $SU(2)_L$ scalar triplet
with $Y=-3$. Since the triplet cannot couple directly with the SM fermions it has no significant impact on other low energy observables. Only in these cases does the model provide a positive shift in 
the $W$ mass, besides an enhancing of $(g-2)_\mu$ value, and other processes~\cite{Crivellin:2023xbu}. It has been noted in Ref.~\cite{Strumia:2022qkt} that, in general, if the shift of the $W$ mass is due to new physics at the tree level, collider bounds can be easily satisfied, while if the new physics solves it only at the loop level it has generic problems that can only be evaded in special situations. This is the price to be paid for \textit{ad hoc} solutions to the $W$ mass anomaly in the context of the SM symmetries.
%It is well known that the SM does not explain at tree level the $W$ and $Z$ %masses and that radiative corrections are needed in order to have agreement %with the experimental data, but this was before the result of the CDF %II~\cite{CDF:2022hxs}. This shift in vector boson masses may occur in models %other than the SM. However, the shift is not arbitrary, for example, it has %been known for a long time that a real singlet produces a shift %\cite{Lopez-Val:2014jva} but it is not able to explain the mass of $W$ %obtained in the CDF II~\cite{Sakurai:2022hwh}. Moreover, as stressed in %Ref.~\cite{Strumia:2022qkt}, new physics that contributes to the $W$ mass at %the loop level must be around the weak energy scale to fit the anomaly, %however, it is generically in conflict with collider bounds. It is for this %reason that a search for a tree level explanation of the $W$ boson mass shift %is warranted~\cite{Evans:2022dgq}.
%This has been shown in the case of two Higgs doublet extensions of the SM, one %of them inert (it does not contribute to the $M_W$ at the tree %level)~\cite{Babu:2022pdn}. 
%Electroweak models with left-right gauge symmetry is an example of the tree %level explanation of the $W$ anomaly. 
Other solutions to the $W$ mass shift have been proposed in the literature. For instance, it may be new physics whose effects appear through radiative corrections~\cite{Strumia:2022qkt,Sakurai:2022hwh,Babu:2022pdn}, or also at tree level~\cite{FileviezPerez:2022lxp,Evans:2022dgq,Senjanovic:2022zwy,Lazarides:2022spe}.

%\textcolor{red}{A characteristic that we will explore here is the fact that in %models with many scalar fields, some of them do not have couplings with the %known fermions.
%Moreover, they may, or may not, contribute to the spontaneous breaking of the symmetries. For instance, if a symmetry, or the representation content, of the 
%model, prohibits a scalar from interacting with fermions but it contributes to %the mass of $W$, the respective VEV would be, in principle, a free parameter %and it would be in charge of the extra contribution to the %$W$'mass.\textbf{este paragrafo em vermelho seria removido.} }

An interesting example is adding only one real ($Y=0$) scalar triplet to the SM which does not couple to any lepton and which contributes only to the $W$ mass which has been considered in Refs.~\cite{FileviezPerez:2022lxp,Evans:2022dgq}. In this case, the vacuum expectation value of the real triplet scalar implies
\begin{equation}
\delta W=2g^2v^2_T,
\label{vt1}
\end{equation} 
being $v_T\approx 3.2$ GeV as required to solve the $W$ mass tension~\cite{Evans:2022dgq}.
Another example of this mechanism occurs in the supersymmetric models in which the supersymmetry forbids the interaction of at least one scalar with fermions but its VEV contributes to the vector boson masses~\cite{Rodriguez:2022wix,Rodriguez:2022hsj}.

Finally, as the result of the CDF II has a precison  not even LEP nor the recent LHC have had, we can assume that the latter results are the correct one and CDF II has discovery new physics effect, in other words, we assume that there is no exist a tension in the $W$ mass measurements but CDF II has really found new physics effect.

The outline of this paper is as follows: in Sec.~\ref{sec:lrs}, we consider two versions of the model with a left-right gauge symmetry given in Eq.~(\ref{lrmodels}). In the first case, considered in Subsec.~\ref{subsec:model1}, we impose, as usual, a generalized parity and all the symmetries are broken spontaneously; In the second case Subsec.~\ref{subsec:explicitpb}, we study a version of the model without a generalized parity symmetry since it is broken explicitly. In both cases the masses of the charged and neutral vector bosons receive extra contributions from the doublet(triplet) $\chi_L$. Our conclusions are in the last section, Sec.~\ref{sec:con}.

\section{Model with left-right symmetry}
\label{sec:lrs}

We will assume that the mass of the $W$ boson is given in Eq.~(\ref{cdf}) and try to understand it in the context of the well-motivated models with electroweak gauge symmetry
\begin{equation}
	SU(2)_L\otimes SU(2)_R\otimes U(1)_{B-L},
	\label{lrmodels}
\end{equation}
with~\cite{Senjanovic:1975rk,Mohapatra:1979ia,Senjanovic:1978ev,Diaz:2019ebd}, or without~\cite{Diaz:2020pwh}, a conserved (at a given energy scale) generalized parity $\cal{P}$.

In this sort of models left- and right-handed fermions transform as doublets of $SU(2)_L$ and $SU(2)_R$, respectively. The minimal scalar sector may consists of one bidoublet $\Phi\sim(\mathbf{2}_L,\mathbf{2}_R,0)$, and two doublets $\chi_L\sim(\mathbf{2}_L,\mathbf{1}_R,+1)$, \cancel{and} $\chi_R\sim(\mathbf{1}_L,\mathbf{2}_R,+1)$. Instead of doublets, these scalars may be complex triplets, say $\chi_L\sim(\mathbf{3}_L,\mathbf{1}_R,+2)$ and  $\chi_R\sim(\mathbf{1}_L,\mathbf{3}_R,+2)$, if we want to have Majorana neutrino~\cite{Mohapatra:1979ia}. 
%\textcolor{red}{However, we will not consider the latter possibility here %because our idea does not depend on it. \textbf{esta frase vai ser omitida.}}

Hence since the  vacuum expectation value (VEV) of $\langle \chi_L\rangle=v_L/\sqrt{2}$   
 contributes only to the mass of the vector bosons it must have an arbitrary value, and it may be even zero $v_L=0$ , in this case, $\chi_L$ is an inert doublet since it does not couple with fermions nor contribute to the vector boson masses. In fact, to allow the doublets $\chi_{L,R}$ to interact with quarks, new quarks and lepton singlets must be introduced as in Ref.~\cite{Hernandez:2021uxx}. The latter case will not be consider here, since we are interested in the minimal matter content of the model. Here we assume that $v_L\not=0$ and arbitrary, to explain the $W$  mass shift in the context of these models. Of course, $\langle \chi_R\rangle=v_R/\sqrt{2}\not=0$ because this VEV is in charge of the mass of the $W_R\approx W_2$, which is a heavy vector boson. 

Here we will consider explicitly only $\chi_{L,R}$ doublets under the respective $SU(2)$ factor. In this case the mass square matrix for the charged vector bosons (CVB) has the following structure in the symmetry basis $(W_L\,W_R)$
\begin{equation}
	M^2_{CVB}=\frac{g^2_Lv^2_R}{4}\left(\begin{array}{cc}
		x+y& -2\,z\,\epsilon \\
		&1+x
	\end{array}
	\right),
	\label{massws}
\end{equation}
where $k$ and $k^\prime$ denote the VEVs of the bi-doublets and we have defined:

\begin{equation}
	x=\frac{K^2}{v^2_R},\;\;K^2=k^2+k^{\prime2},\;\;y=\frac{v_L^2}{v_R^2},\;\;\;\;z=\frac{kk^\prime}{v^2_R},\;\;\epsilon=\frac{g^2_R}{g^2_L}.
	\label{defpara}   
\end{equation}

We call the mass eigenstates of the matrix in (\ref{massws}) just $W_1$ and $W_2$, similarly with  $Z_1$ and $Z_2$, but we omit the mass matrix of the neutral vector bosons. Notice that in (\ref{massws}) when $z=0$ ($v_R\to\infty$) and $v_L=0$ we have the lighter $W$ and it is exactly the SM's vector boson, being the other one the heavier, which decouples. 

If $\chi_L$ is a doublet we may think that it will contribute with all the other neutral scalars, to the mass of the $W$ as follows $K^2+v^2_L=(246\, \textrm{GeV})^2$. This would happen in any Higgs doublet model in which their VEVs contribute to some fermion masses. However, in the present sort of model $v_L$ does not contribute to any fermion masses and it may be even zero without affecting too much the phenomenology of the model, as in fact, it has been used in most phenomenological studies. Hence, it is a completely free parameter. Hence we can  saturate the mass of the $W$ obtained in the context of the SM with $K^2=(246\, \textrm{GeV})^2$.

\subsection{Model with spontaneously broken parity}
\label{subsec:model1}

In the first model, there is also the symmetry under a generalized parity, $\cal{P}$. In this case, the representation content is determined in order that this parity is conserved before the spontaneousy breaking of symmetry (SBS) and also in Eq.~(\ref{defpara}) with $\varepsilon=1$
and under $\cal{P}$ we have
\begin{eqnarray}
	&&	g_L\leftrightarrow g_R ,\;W_{L\mu}\leftrightarrow W^\mu_{R},\; f_L\leftrightarrow f_R,\nonumber \\&&
	\chi_L\leftrightarrow \chi_R,
	\Phi_i\leftrightarrow \Phi^\dagger_i,\tilde{\Phi}_i\leftrightarrow \tilde{\Phi}^\dagger_i,
	\label{lr}
\end{eqnarray}
With $\;\tilde{\Phi}_i=\tau_2\Phi^*_i\tau_2$, being $\tau_2$ a Pauli matrix~\cite{Senjanovic:1975rk,Mohapatra:1979ia,Senjanovic:1978ev}. After the SBS the parity $\mathcal{P}$ is also broken.

It is important to notice that, the introduction of the doublet (or triplet), $\chi_L$, is only required for keeping a generalized parity $\mathcal{P}$ in Eq.~(\ref{lr}), the doublet does not couple  to quarks or charged leptons, and the triplet is coupled with leptons only if it carries $U(1)_{B-L}$ charge, however, a discrete symmetry, say $\mathbb{Z}_2$, may be introduced in order to forbid the coupling with the triplet $\chi_L$. In this case, there are no extra constraints  on the vacuum expectation value of $\chi_L$ coming from the equivalent to the $\rho$ parameter in the standard model ($\rho=M_{W_1}/c_WM_{Z_1}$).

%As we said above, the scalar doublet, $\chi_R$, under $SU(2)_R$ has a VEV %whose value define the LR symmetry breaking scale, but the $\chi_L$ which is %introduced just to maintain the generalized parity symmetry has a free VEV, %that would be zero, or having an appropriate value to solve the case of $W$ %mass as we will do here. 

In this case, we have the parameters $x,y$ and $z$ defined as in the matrix in Eq.~(\ref{massws}) but now with $\epsilon=1$. The eigenvalues of this mass matrix, using $g_L=g_R$, are: 
\begin{eqnarray}
&& M^2_{W_1}=\frac{g^2v^2_R}{4}\left(x+\frac{1+y}{2}-\sqrt{\Delta_1}\right),
\nonumber \\ &&
M^2_{W_2}=\frac{g^2v^2_R}{4}\left(x+\frac{1+y}{2}+\sqrt{\Delta_1}\right),\nonumber \\&& \Delta_1=4z^2+\frac{1}{4}(1-y)^2,
\label{massws2}
\end{eqnarray}

Expanding in $x$ around $x=0$ $(v^2_R\gg K^2)$ we have
\begin{eqnarray}
M^2_{W_1}&\approx&  \frac{g^2K^2}{4} \left[ 1+\xi-x\left( 1+y+y^2\right)+x^3\right]+\mathcal{O}(x^4),\nonumber \\  M^2_{W_2}&\approx& \frac{g^2}{4}v^2_R+\mathcal{O}(x^2), \quad \xi=v^2_L/K^2.
\label{delta2}
\end{eqnarray}

Notice that $M^2_{W_2}\gg M^2_{W_1}$, and we identify, approximately, the $W^\pm_1$ with the $W^\pm$ of the standard model field.

We can see that the extra mass contributions come from the $v_L$, from Eq.~(\ref{delta2}), we have a $W$ mass shift as in Eq.~(\ref{vt1})~\cite{Evans:2022dgq},
\begin{equation}
\delta M^2_W=\frac{g^2}{4}v^2_L.
\label{deltaMW}
\end{equation}

We use the exact expression in Eq.~(\ref{massws2}) for graphing $M_W$ as a function of $v_L$, at several values for $v_R$, say from 1 TeV to $\infty$ (in fact $v_R\gg k,k^\prime$), using $K^2=(246\,\textrm{GeV})^2$. The result is shown in Fig.~\ref{fig:modelo1}, where the colored strip indicates the value of the CDF-II at the level of $1\sigma$   . We have verified that with $2.4\, \textrm{GeV}\leq v_L\leq 60.6$ GeV, the lower bound refers to $v_R\to \infty$ and the larger one to $v_R=1$ TeV, we get the mass of the $W$ obtained by the CDF~\cite{CDF:2022hxs}. Also, we will see that the solution with $v_R=24$ TeV, and $v_L=3.47$ GeV is another possibility for explaining the $W$ mass shift. For the sake of simplicity, we have assumed $k=k^\prime$. Moreover, we use $g=0.6539$ and $s_\theta=0.2312$~\cite{Workman:2022ynf}. Of course, although the model has two mixing angles in the neutral vector bosons \cite{Kokado:2015iya,Diaz:2019ebd}, in the limit of the large mass of $W_2$ ($v_R\to\infty$) we can identify
$\sin\theta = \sin\theta_W$, where $\theta_W$ is the electroweak mixing angle of the SM.

\begin{center}
	\begin{figure}[!h]
		\includegraphics[width=9cm]{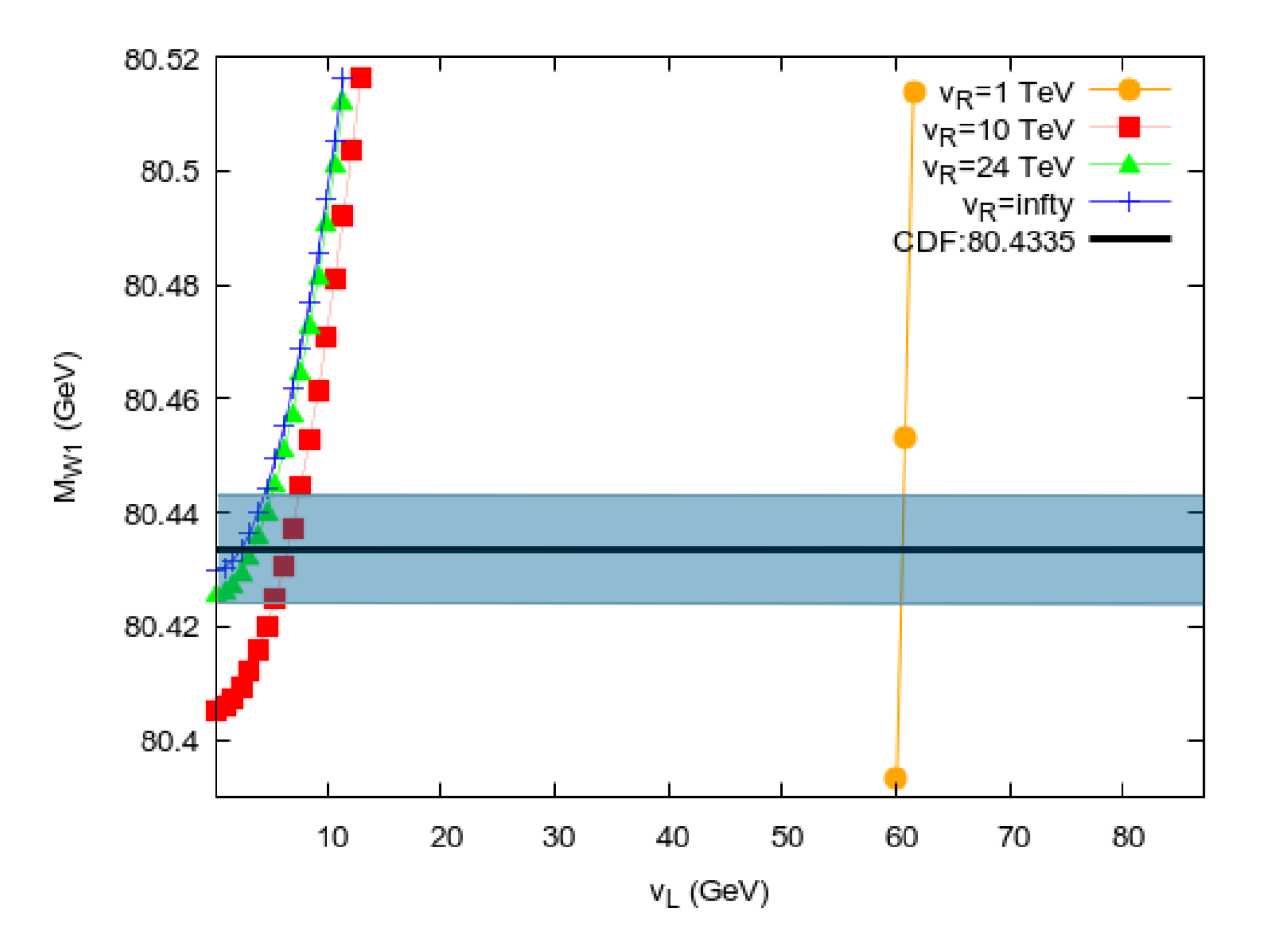}
		\caption{$W$ mass as a function of $v_L$ in a model with left-right symmetry at many values of $ v_R$. }
		\label{fig:modelo1} %figx
	\end{figure}
\end{center}	

However, there are also contributions to the $Z_1$ and $Z_2$ masses (we omit the mass square matrix in the neutral vector bosons) given by
\begin{eqnarray}\label{EN1}
M^2_{Z_1}\,&=&\,\frac{g^2\,v^2_R}{4}\left(x+\frac{1}{2}(1+y)(1+\delta^\prime)-\frac{1}{2}\sqrt{\Xi_1}\right),\nonumber \\
M^2_{Z_2}\,&=&\,\frac{g^2\,v^2_R}{4}\left(x+\frac{1}{2}(1+y)(1+\delta^\prime)+\frac{1}{2}\sqrt{\Xi_1}\right),
\end{eqnarray}
where:
$$\Xi_1\,=\,(1+y)^2(1+\delta^\prime)^2-4\,\delta^\prime
x(1+y)-4(1+2\delta^\prime)y+4x^2$$
If we denote $g^\prime$ the gauge coupling of the $U(1)_{B-L}$ factor, then $\delta^\prime=g^{\prime \;2}/g^2=\tan^2\theta=0.3987$ and expanding in $x$
\begin{equation}
M^2_{Z}\approx \frac{g^2K^2}{4\cos^2\theta}(1+\xi),\;\;\xi=v^2_L/K^2
\label{massz1},
\end{equation}
however the usual relation
\begin{equation}
\delta M^2_Z\approx \frac{\delta M^2_W}{\cos^2\theta},
\label{massz2}
\end{equation}
is still valid, as expected, since $\chi_{L,R}$ is a doublet of the respective $SU(2)$ factor, then still $\rho=1$ at tree level. If $\chi_L$ is a complex triplet, i.e., carrying hypercharge, the relation in Eq.~(\ref{massz2}) implies an upper bound in its VEV, denoted by $v_T$, in order to keep $\rho=1$, within the experimental error, then $v_T\stackrel{<}{\sim} 4$ GeV which is also consistent with the CDF II result of the $W$ mass. This happens in any model with this sort of scalar~\cite{Montero:1999mc}. Recently, it has been shown that, in fact, a scalar triplet can explain the $W$ mass anomaly but in a region of the space parameter that has already been excluded by measurement of the standard model final states, mainly those with four charged leptons~\cite{Butterworth:2022dkt}. 

From Eq.~(\ref{massz2}), $\delta M_W=0.0094$ GeV implies that $\delta M_Z=0.0107$ GeV. Thus, in this class of models, the mass of the neutral bosons which is equivalent to the SM's $Z$, must be also different than its value¸ obtained in the context of the SM. Since this is a prediction of our model, it makes it falsifiable. 

\subsection{Model with explicit parity breaking}
\label{subsec:explicitpb}

Let us now consider a model with the gauge symmetry as in Eq.~(\ref{lrmodels}) but with $\cal{P}$ being explicitly broken~\cite{Diaz:2020pwh}, i.e., in which $g_L\not=g_R$ and the transformations in Eq.~(\ref{lr}) are not valid. Hence, the bosons and all parameters in the $SU(2)_L$ are not related to those of the $SU(2)_R$ factor, including the mixing matrices in each chiral fermion sector. In this case even it is not necessary to introduce any doublet or triplet under the $SU(2)_L$ factor.

 Moreover, in this case, since the generalized parity, $\mathcal{P}$, is broken explicitly, there is no transformations as those in Eq.~(\ref{lr}) at all, then a real triplet under the $SU(2)_L$ factor i.e., $\chi_L\sim(\mathbf{1}_R,\mathbf{3}_L,0)$, may be introduced and since it this triplet does not carry $U(1)_Y$ charge it does not contribute to the $Z$ mass neither to neutrino (Majorana) masses. See Refs.~ \cite{FileviezPerez:2022lxp,Evans:2022dgq} for details.

Here, however we will continue to use a doublet scalar under the $SU(2)_L$ factor, $\chi_L$. In this case, the mass matrix of the charged vector bosons is as in Eq.~(\ref{massws}) with $x,y$, and $z$ defined as in the previous model but now $\epsilon =g^2_R/g^2_L\not=1$.
Then we obtain for the charged vector bosons the masses:
\begin{eqnarray}
M^2_{W_1}&=&
\frac{g^2_L\,v^2_R}{8}\left[(1+\epsilon)\,x+y+\epsilon-\sqrt{\Delta_2}\right]
\nonumber \\ M^2_{W_2}&=&
\frac{g^2_L\,v^2_R}{8}\left[(1+\epsilon)\,x+y+\epsilon+\sqrt{\Delta_2}\right], \nonumber \\
\Delta_2&=&4\epsilon x^2+[\,\epsilon-y-(1-\epsilon)\,x]^2,
\label{masssm2}
\end{eqnarray}
from these equations we see that
\begin{eqnarray}
\delta M^2_{W_1}&\approx& \frac{g^2_L}{4}K^2\left[\xi-\left(1+\frac{y}{\epsilon}\right)x+\left(1-\frac{y}{\epsilon}\right)x^2\right]+\mathcal{O}(x^3)\nonumber \\ &=&\delta M^2_{W_1}\vert_{\epsilon=1}+\delta M^2_{W}
\vert_{\epsilon\not=1}+\mathcal{O}(x^3)\nonumber \\
M^2_{W_2}&\approx& \frac{g^2_L}{4}\,\epsilon\, v^2_R(1+x)+\mathcal{O}(x^2)=\frac{g^2_R}{4}v^2_R(1+x)+
\mathcal{O}(x^2),
\end{eqnarray}
with $\;\delta M^2_{W}\vert_{\epsilon=1}$ is as in Eq.~(\ref{deltaMW}) but now $g\to g_L$, where we have an extra contribution
\begin{equation}
\delta M^2_{W}\vert_{\epsilon\not=1}=-	\frac{g^2_L}{4}K^2\left[\left(1+\frac{y}{\epsilon}\right)x-\left(1-\frac{y}{\epsilon}\right)x^2\right],
\label{deltaMW2}
\end{equation}
and $\delta M^2_{W}\vert_{\epsilon\not=1}\to 0$ when $x\to 0$. Hence the $\delta M^2_W$ is again as given in Eq.~(\ref{deltaMW}).

Notice that, in order to maintain the $\delta M_{W}$ mass a certain value,
the greater the $v_R$ ($x\ll1$) smaller the $\epsilon\ll1$ ($g_R\ll g_L$) at a given energy. This also keep the mass of the $W_2$ at a given value.
In Fig.~\ref{fig:modelo2}(a) we show the $W$ mass in terms of $\epsilon$ for several values $v_L$ and $v_R=1$ TeV. We see that even with $v_L=0$ it is possible to accommodate the value of the $W$ mass obtained by CDF II. In Fig.~\ref{fig:modelo2}(b) we show the mass of $W$ in terms of $v_L$ for several values of $\epsilon$ and also $v_R=1$ TeV. 

For neutral vector bosons, we have in this case
\begin{eqnarray}\label{GBN5}
M^2_{Z_1}&=&\frac{1}{8}g^2_Lv^2_R\left(x (1+\varepsilon)+y\,(1+\delta_L)+\,\varepsilon+\delta_L- \sqrt{\Xi_2}\right)\nonumber \\
M^2_{Z_2}&=&\frac{1}{8}g^2_Lv^2_R\left(x (1+\varepsilon)+y\,(1+\delta_L)+\,\varepsilon+\delta_L+\sqrt{\Xi_2}\right),
\nonumber \\
\Xi_2&=&y^2(1+\delta_L)^2+x^2(\varepsilon+1)^2+(\delta_L+\varepsilon)^2+
%\nonumber\\&+&
4xy(1-\varepsilon(1+\delta_L)-\delta_L),\;
\end{eqnarray}
where $\delta_L=g^{\prime\, 2}/g^2_L=0.3987$. Again when $x,y\to 0\,(v_R\to\infty)$ the relation in Eq.~(\ref{massz2}) is obtained, independently of the value of $\epsilon$.
\begin{center}
	\begin{figure}[!h]
		\includegraphics[width=9cm]{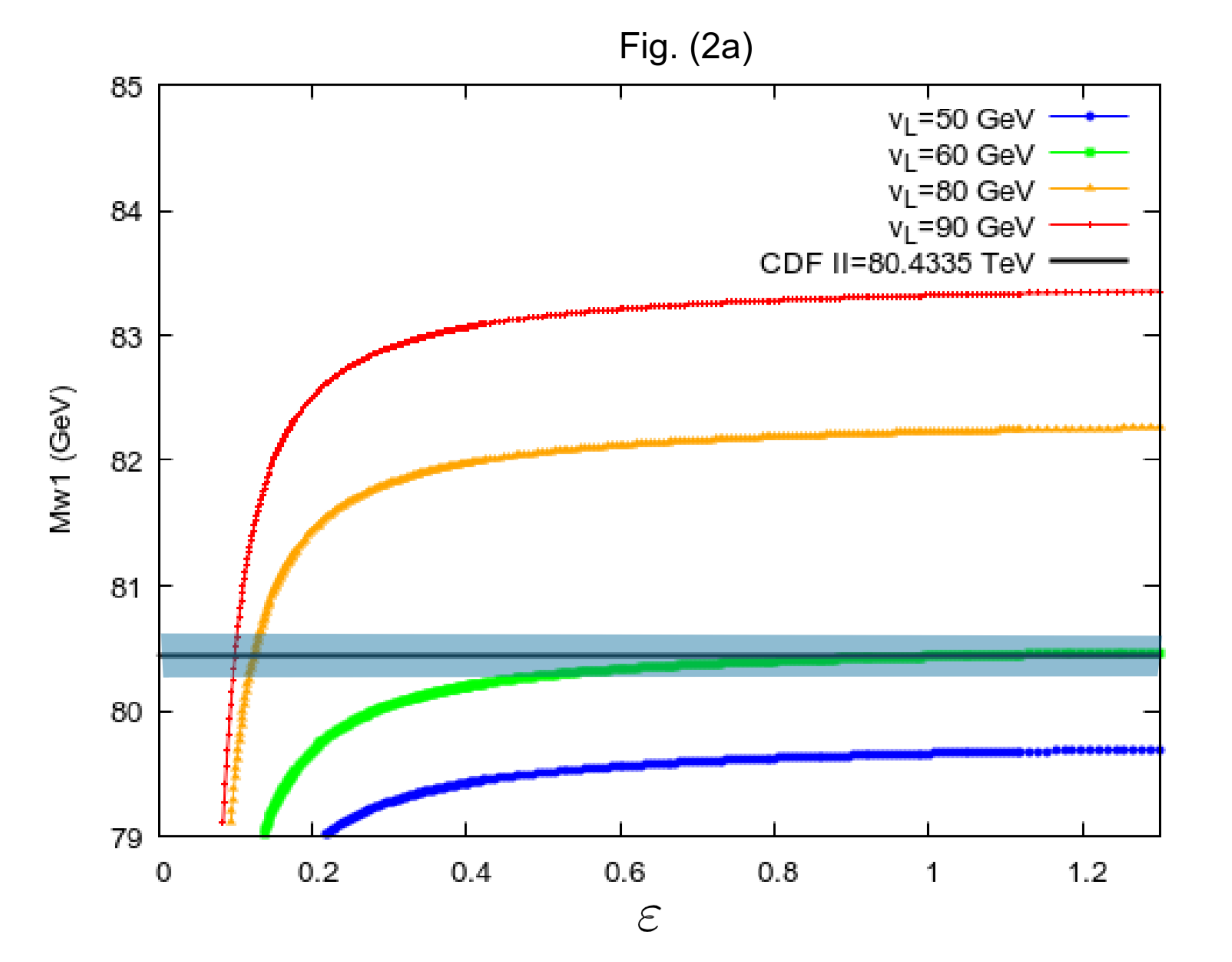}
		\includegraphics[width=9cm]{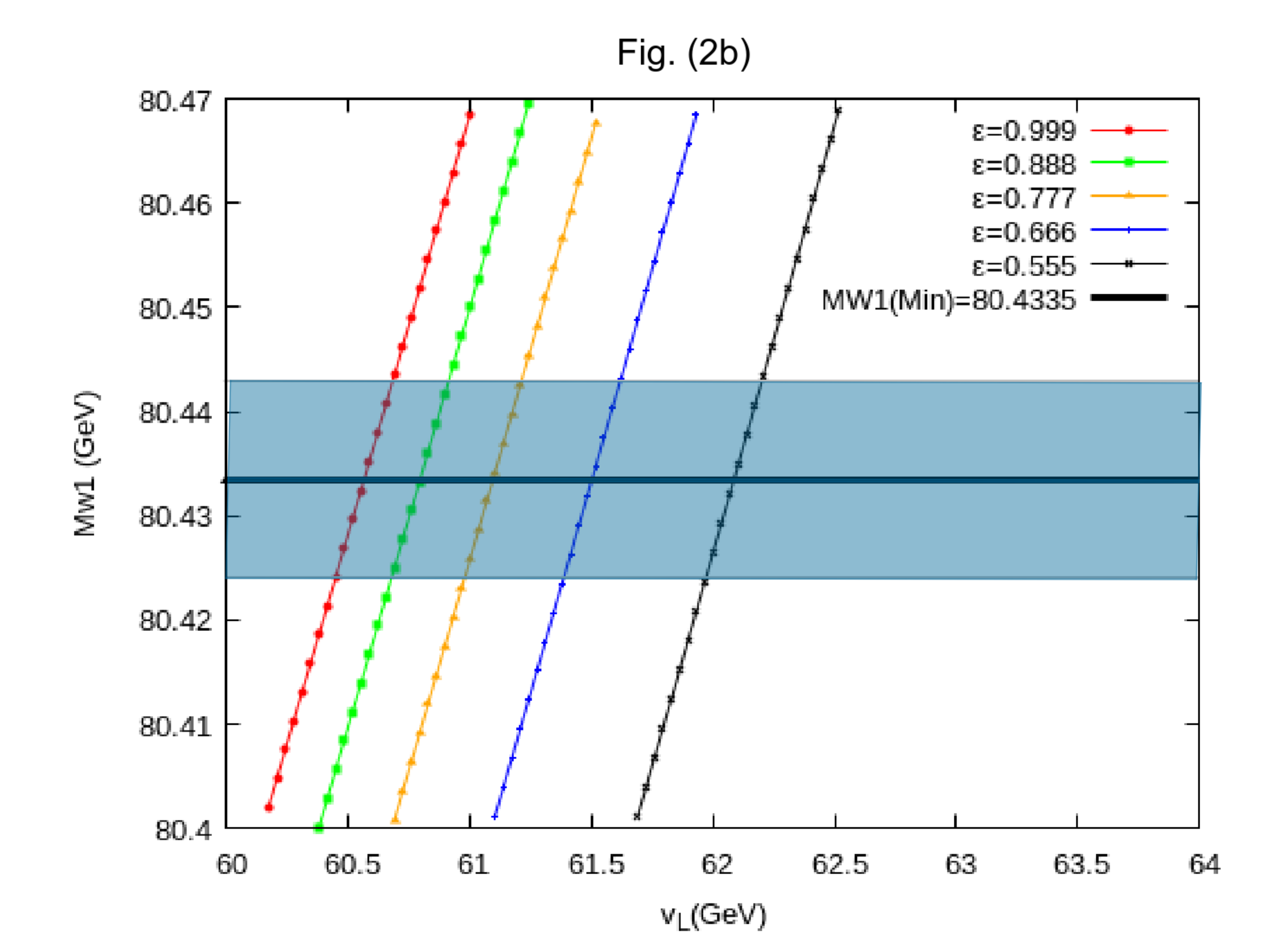}
		\caption{In Fig.~(2a) we show the $W$ mass as a function of $\epsilon$ and several values of $v_L$. In Fig.~(2b) we show the $W$ mass as a function of $v_L$ for several values of $\epsilon$. In both cases $v_R= 1$ TeV.}
		\label{fig:modelo2} %figx
	\end{figure}
\end{center}	
Most searches for $W_2$, and also for $Z_2$, usually denoted by $W^\prime$ and $Z^\prime$, respectively, assume that the couplings of these vector bosons to quarks and leptons are taken to be identical with those of SM's $W$ and $Z$, and also that $g_L=g_R$~\cite{Workman:2022ynf}. This is not the case in the present model which has explicit parity breaking and in this case, the limits are considerably weaker than for the special case of  manifest left-right symmetry. Moreover, since we have considered only Dirac neutrinos and no scalar triplets (no doubly charged scalars), the results obtained in Ref.~\cite{Frank:2018ifw} do not apply straightforward, moreover in the model with explicit parity breaking does not exist any relation between the left- and right- parameters, $g_L$, $g_R$ and the respective mixing matrices at any energy scale.

In particular, in Fig~\ref{fig:modelo2}(a) for $v_R=1$ TeV and $v_L=60$ GeV, we obtain $0.8 \stackrel{<}{\sim}\epsilon \stackrel{<}{\sim}1.2$. On the other hand, we have found but do not show it in Fig.~\ref{fig:modelo2}(a), that when $v_L=0$ the solution still exists but only at the $5\sigma$ level with $\epsilon$ rather small, $g_R\sim10^{-6}g_L$. Then in this case the right-handed currents are superweak.

\vspace{4cm}

\section{Conclusions}
\label{sec:con}

As we said in the Introduction, it would be strange that the masses of the $W$ and $Z$ bosons, measured by assuming   the SM particles as input, these remain the same when new physics is discovered. Deviations would appear if there are extra contributions to $W$ and $Z$ decays. Thus, measurements with a greater precision will, necessarily, show deviations from any of the SM predictions, if new physics is around the corner, they would appear even with a lower precision if we know where, and how, to look.

In the present models unlike all previous proposals, we use only the minimal version of the well motivated models with $SU(2)_L\otimes SU(2)_R\otimes U(1)_{B-L}$ symmetry. 

Moreover, the vacuum expectation value $v_L$, which is usually considered equal to zero, allows to accommodate the value of the $W$ obtained by the CDF II, if $v_L\sim$ 4 GeV and $ v_R=1$ TeV, with this value, it certainly will not have a significant impact on other phenomenological processes but the vector boson masses. It is interesting that if these models are the solution to the $W$ mass, there must also be an extra contribution to the mass of $Z$. In the first model, if $v_L=0$ there is no solution (at least at tree level) to the $W$ mass shift  and in this case $\chi_L$ is just an inert doublet, while the second model even with $v_L=0$ still has a solution for the tension of the $W$ mass and $\chi_L$ is again an inert doublet with the right-handed currents being superweak, see Sec~\ref{sec:lrs}. Hence, in both models $v_L\not=0$ is a better option. 

%In fact, the solutions for the $W$ mass anomaly depend on the value of $v_L$ %and $v_R$, for instance, $v_R=24$ TeV, and $v_L=3.47$ GeV is an interesting %possibility as we have pointed out before. Models with left-right symmetry %have been well studied, even with $v_L=0$, and with such a small value for %$v_L$ GeV, their phenomenological consequences, for instance, weak decays, %muon $g-2$ factor, etc, will not be significantly altered. 

Models with the same left-right symmetry as in the present work have been considered to resolve the displacement of the $W$ mass, however, these authors resort to non-minimal versions of the representation content of the model, by adding, to the minimal representation content, vector-like quarks for they can solve the usually called $W$ mass anomaly~\cite{Dcruz:2022rjg}.  Moreover, unlike this version of the model, here we do not assume that the vacuum expectation value of $\chi_L$ is the main responsible for the spontaneous symmetry breaking of the SM gauge symmetry, we leave this issue for the VEVs in the bidoublet. It is more interesting to see  the possibility that the minimal representation content of any model can solve the issue of the positive shift of the $W$ mass. We have done this in models with the gauge symmetries \textcolor{red}{}in Eq.~(\ref{lrmodels}).

%It is worth emphasizing that, as we are considering a complete ultraviolet %renormalizable model it is not necessary, at least at the beginning, to %consider radioactive corrections, oblique parameters as $T$, $S$, and $U$ or, %to consider an effective theory.  

%\textbf{Moreover, in the context of the SM the $W$ depends on the imput %parameter as $M_Z,\alpha, G_F$ and also on the Higgs mass $m_H$. It seems %clear that in any extension of the standard model new inputs have to be %intruducesd and, the $W$ mass will be different from the SM vaue at some level %data's precision. \textcolor{red}{\textbf{este paragrafo pode ser eliminado.}}}

%If our prediction of the change in the mass of the SM's $Z$, $\delta %M_Z=0.0107$ GeV, is experimentally confirmed and, from the results of %Ref.~\cite{Butterworth:2022dkt} are valid for the triplets $\chi_L$ in %left-right symmetric models, this would favor that neutrinos being Dirac %particles.
\newpage

%\section*{acknowledgments}
\noindent{\textbf{Acknowledgments} }\\
H.D. and O.P.R. are thankful for the full support and V.P. for the partial support of Universidad Nacional de Ingenieria funding Grant FC-PFR-45-2023.

\end{document}